# The Dark Photon in the PHELEX Experiment

A.V. Kopylov, I.V. Orekhov, V.V. Petukhov, and A.E. Solomatin
Institute for Nuclear Research of RAS, 117312 Moscow, Prospect of 60th Anniversary of October Revolution 7A, Russian Federation

This paper presents diurnal measurements of the single-electron count rate in the PHELEX experiment using multi-cathode counters over the period from 2023 to 2026. The results obtained are interpreted as an indication of a possible effect from dark photons with masses ranging from 10 to 40 eV. The correlation of the measured rates in counters with Fe and Al cathodes suggests that the effect does not depend on the magnetic properties of the cathode material. This can be interpreted as an argument against many dark matter models related to magnetic moments. The result, if confirmed in an independent experiment, could narrow the range of possible candidates to those that interact primarily through an electric (or longitudinal) field, insensitive to spin. A dark photon with kinetic mixing fits this description ideally.

## 1. Introduction.

Numerous attempts to detect an effect from dark matter in one form or another have not yielded the expected result, which is a significant discouraging factor for modern experimental physics. Under these circumstances, it is extremely important to utilize any available instrumentation where there is even the slightest chance of obtaining a positive result. In the PHELEX (PHoton-ELectron EXperiment) experiment, we have undertaken an attempt to "see" the effect from dark photons as a dark matter candidate. The first publications on hidden photons date back to the 1980s [1-3]. In our experiment, we aim to register the effect from dark photons through diurnal variations in the single-electron count rate using a gas proportional counter with a large (approximately 30 dm²) metal cathode surface in a multi-cathode configuration to achieve a sufficiently high (> $10^5$) gas gain factor. In this method, it is assumed that the dark photon (4-vector $A'$) mixes with the ordinary photon (4-vector $A$), which in turn causes the photoelectric effect on the metal cathode surface, followed by the emission of a single electron [4]. The target in this method is the free electrons of the degenerate electron gas of the metal. The count rate is determined by the cathode area, the kinetic mixing constant, and the quantum efficiency of the cathode. We considered the dark photon mass range from 10 to 40 eV, as this mass corresponds to the vacuum ultraviolet energy range where metals have the highest quantum efficiency. At such small masses, we obtain a high occupancy of dark photons. For an average dark matter density in the solar vicinity of 0.4 GeV $cm^{-3}$ [5], we obtain a dark photon occupancy $n$ from $1\times10^7$ to $4\times10^7$ $cm^{-3}$. At a dark photon velocity $v \sim 220$ km/s = $2.2\cdot10^7$ cm $s^{-1}$, the dark photon flux for a mass of 10 eV is: $F = nv \approx 8.8\cdot10^{14}$ $cm^{-2}$ $s^{-1}$. This is a huge flux, much larger than that of solar neutrinos. In the PHELEX experiment, we focus on searching for diurnal variations in the count rate in sidereal days, as the gold standard for their astrophysical origin, and the absence of such variations in solar-terrestrial days, as the null hypothesis. Such variations are possible during Earth's rotation if the Sun occasionally crosses a region with a certain alignment of dark photons relative to the Earth's axis. The effect should be observed when the longitudinal component of the dark photons is oriented at a specific angle relative to the Earth's axis. In the PHELEX experiment, we conducted measurements from January 2023 to February 2026. During this time, 16 Runs were carried out, each Run lasting approximately 60 days. The last 8 Runs were performed with two counters: with Fe and Al cathodes, to determine whether the magnetic properties of the cathode material influence the effect. The measurement results and their interpretation are presented below.

**2. Measurement Results.**

The design of the multi-cathode counter and the method are described in detail by us in [6, 7]. A cylindrical gas proportional counter with a volume of approximately 15 liters, filled with a mixture of Ne + 10% $CH_4$ at a pressure of 0.1 MPa, is used. The counter has a cylindrical metal cathode with an area of approximately 30 $dm^2$. The anode is a gold-plated W-Re alloy wire with a diameter of 20 μm stretched along the axis of the cylindrical cathode. To ensure a high ($>10^5$) gas gain factor, a second cathode made of 20 stretched nichrome wires with a diameter of 50 μm is positioned at a radius of 20 mm from the anode. Electrons emitted from the cathode surface drift towards the second cathode and then to the anode wire, near which avalanches with multiple electron generation develop. The signal from the anode wire is fed to the input of a charge-sensitive preamplifier, and from its output to the input of an analog-to-digital converter board. The signal is digitized at a frequency of 10 MHz. With a pulse width of average amplitude from a single electron of approximately 100 μs (at half its height), this allows tracking the pulse shape with subsequent noise discrimination in offline mode. Approximately 1 Tb of data is accumulated per day of measurements. The results obtained in the first 12 Runs were published in [8]. In the first series of 4 Runs from January to July 2023, we observed an excess above the average count rate in all four Runs during sidereal days in the 4-hour interval from 8:00 to 12:00. The significance of the noted anomaly was $>6\sigma$ [8]. In the second series, also of 4 Runs from October 2023 to March 2024, we observed a similar anomaly in all four Runs, also in a 4-hour interval, but during the period from 18:00 to 22:00 in sidereal days. The significance of this anomaly was $>4\sigma$ [8]. In both series of measurements in Runs 1 to 8, no similar anomalies were observed in solar-terrestrial days in any of the 4-hour intervals throughout the day [8], which ruled out the possibility of influence from any man-made processes on Earth or caused by sudden solar activity on the results. The measurement results in subsequent Runs 9 to 15, already with two detectors, with Fe and Al cathodes, did not reveal similar anomalies, which is evidence that the results obtained in the first two series are not an instrumental effect, are not caused by processes on Earth or the Sun, and are quite reliable. In Run 16, for January and February 2026, with two detectors, we again obtained an interesting result. Figure 1 shows the measurement results for this Run.

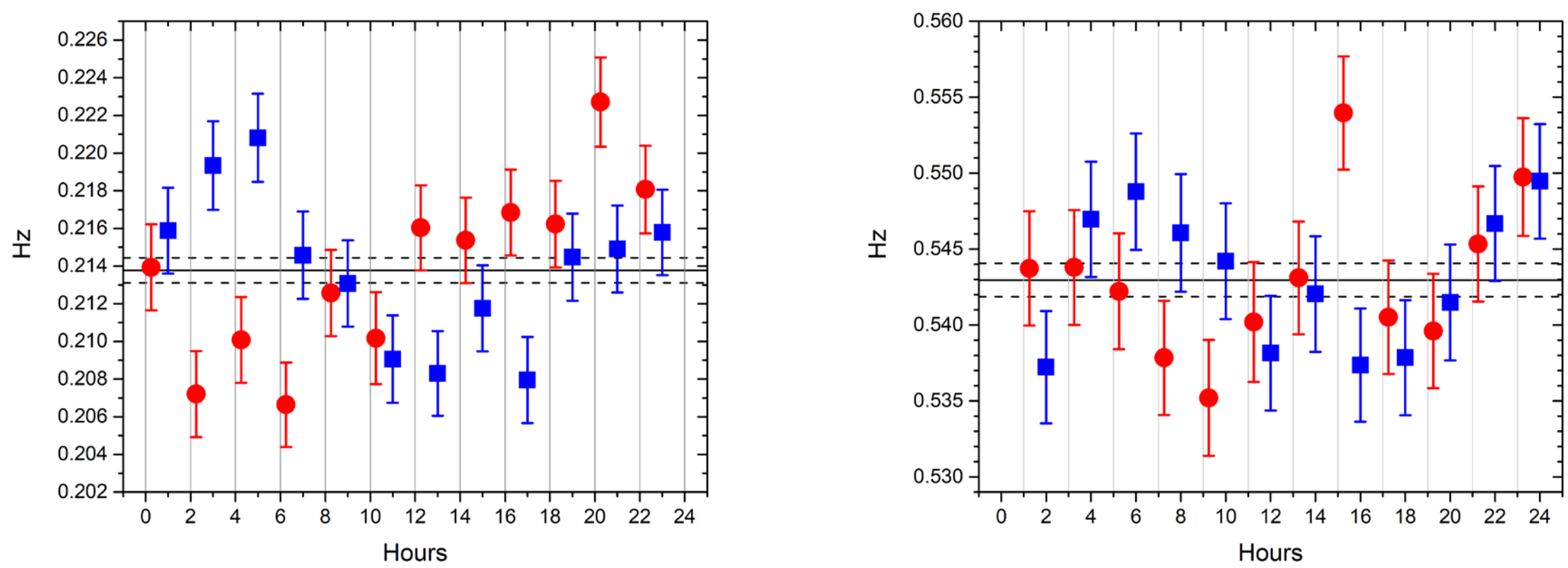

Figure 1. Diurnal variation of the single-electron count rate. Left – for the Fe counter, right – for the Al counter. Squares – in sidereal days, circles – in solar-terrestrial days.

The figure shows that in Run 16, a synchronous change in the count rates for the Fe and Al counters is observed in sidereal days. At the same time, for approximately one half of the sidereal day, the count rate exceeds the average value, and for the second half of the sidereal day, it is below the average value, as expected from our calculations for an angle of the ***E*** vector relative to the Earth's rotation axis within approximately 20° to 50° for our counters, located along the meridian in the laboratory in Moscow [9]. It should also be noted that, as before, no such phenomena were observed in solar-terrestrial days this time either. This is the third time we have observed an anomaly, and each time precisely in sidereal days. In solar-terrestrial days – no anomalies.

As seen in Fig. 1, the observed effect is approximately the same for the Fe and Al counters. This can be interpreted as meaning that the energy transfer mechanism does not depend on the electron spin or the magnetization of the cathode. The result, if confirmed in an independent experiment, could narrow the range of possible candidates to those that interact primarily through an electric (or longitudinal) field, insensitive to spin. A dark photon with kinetic mixing fits this description ideally.

Over the 60 days of the Run, the Sun travels approximately 8 AU along its trajectory. The question arises: were the correlations we observe in Fig. 1 for the Fe and Al counters accumulated throughout the entire path of the Sun, or were they accumulated over some segment of this path? To clarify this issue, we processed the data separately for January and February of Run 16. The results of the measurement data processing are shown in Fig. 2.

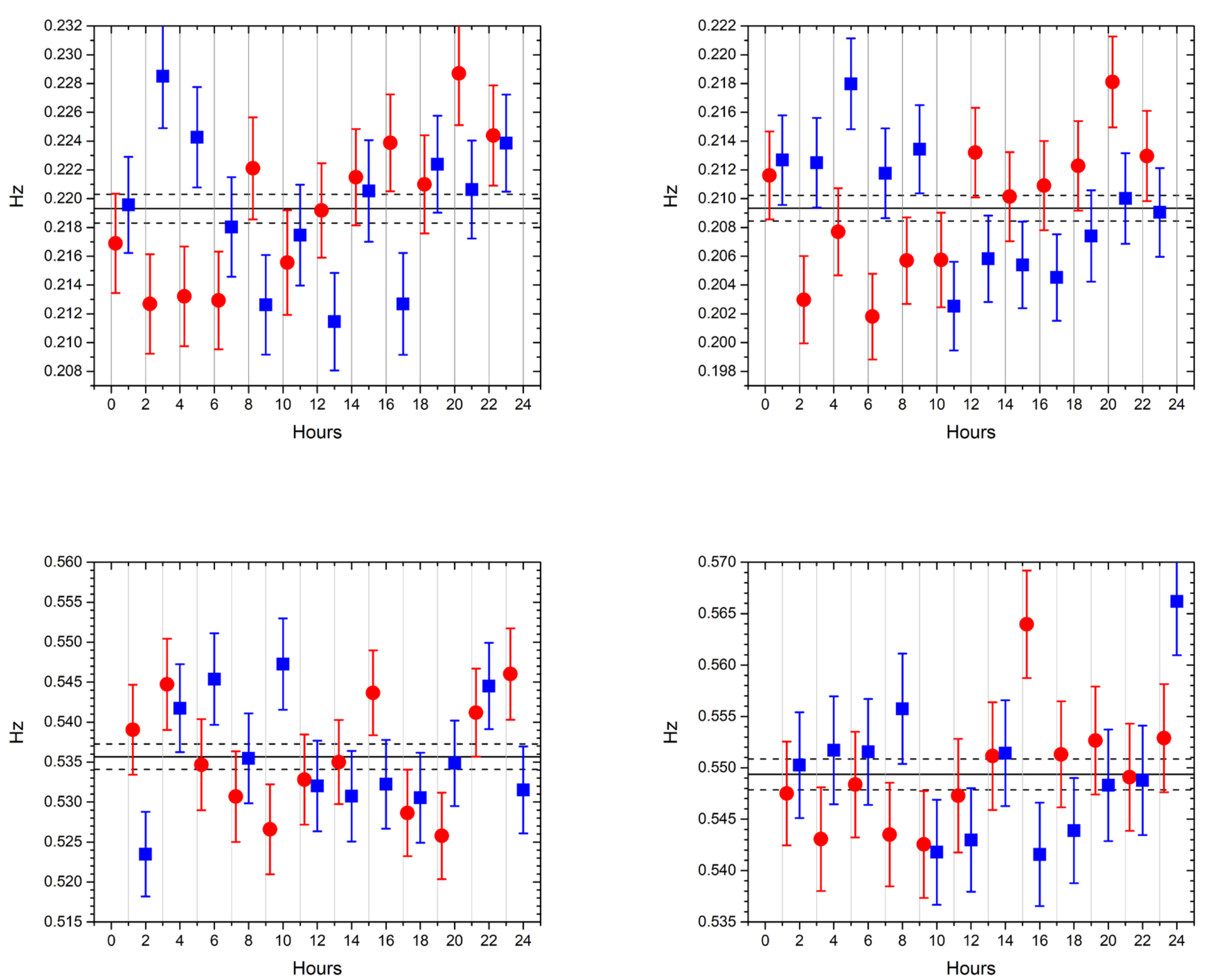

Figure 2. The upper two figures refer to the Fe counter for January (left) and February (right), the lower ones – for the Al counter, respectively. Notations are the same as in Fig. 1.

As seen in Fig. 2, the effect was accumulated continuously over the two months of this Run, and the statistics were accumulated roughly evenly. It follows that the size of the region with aligned dark photons exceeded 8 AU, assuming the Sun crossed a region tied to galactic coordinates. If the region with alignment moved along its own trajectory intersecting the Sun's trajectory, then this option must be considered separately. These results also show that the observed effect is not a product of some sudden electromagnetic interference of terrestrial origin or short-term flare activity on the Sun. It should also be mentioned that in our experiment, both detectors were shielded by a layer of steel passive shielding 40 cm thick, and pulses from all ionizing particles crossing the counter's active volume were discriminated by pulse amplitude. This practically excludes the possibility of an effect through any channel of single-electron generation by deeply penetrating ionizing particles of galactic cosmic rays. It should be emphasized once again that the effect is not observed continuously: in 7 Runs from 9 to 15, from October 2024 to December 2025, i.e., for over a year, the effect was not observed. The source of this effect is not related to terrestrial or solar processes; it is clearly of galactic origin.

Table 1. Data for Runs 1 to 16.

| Run | Date | Cathode | Anomalies | |
|---|---|---|---|---|
| | | | Sidereal Days | Solar-Terrestrial Days |
| 1 – 4 | I/2023 – VII/2023 | Fe | 08:00 – 12:00 (>6 σ) | None |
| 5 – 8 | X/2023 – III/2024 | Fe | 18:00 – 22:00 (>4 σ) | None |
| 9 – 15 | X/2024 – XII/2025 | Fe+Al | None | None |
| 16 | I/2026 – II/2026 | Fe+Al | 3:00 –7:00 (≈ 4σ) | None |

Table 1 presents data for all 16 Runs. In Run 16, a correlation of count rates in the Fe and Al counters is observed throughout the sidereal day. In solar-terrestrial days – no correlations are observed.

**4. Conclusion.**

The measurement results for 2023–2026 have convincingly demonstrated that an effect of diurnal variations in the single-electron pulse count rate is observed in sidereal days. In solar-terrestrial days, we observe nothing similar, as seen in Table 1. This can be interpreted as an effect from the conversion of dark photons into photoelectrons on the surface of the counter cathode. The effect does not depend on the magnetic properties of the cathode material. This suggests that the energy transfer mechanism from dark photons to the free electrons of the metal does not depend on the electron spin or the magnetization of the cathode. The result, if confirmed in an independent experiment, could narrow the range of possible candidates to those that interact primarily through an electric (or longitudinal) field, insensitive to spin. A dark photon with kinetic mixing fits this description ideally. Diurnal variations in the count rate are the gold standard proving the astrophysical origin of the effect. To obtain a strong argument in favor of this hypothesis, it would be useful to place one counter in a position perpendicular to the one currently used. The diurnal variation pattern depends on the detector orientation [9]. However, here we encounter a "conflict of interest": since there are only two counters and a third is not expected in the near future for many reasons, we must choose: use both in the same position in a coincidence mode to ensure reliability or move one to a position perpendicular to the current one in pursuit of a strong "for or against" argument. It would also be very useful to conduct independent measurements in another laboratory at geographic latitudes close to Moscow (Novosibirsk or Tomsk). The phase difference in diurnal variations corresponding to the different geographic longitudes of the laboratories would become a decisive argument in this matter.

**References.**